\documentclass[conference]{IEEEtran}
\IEEEoverridecommandlockouts
\usepackage{cite}
\usepackage{balance}
\usepackage[pdftex]{graphicx}
\usepackage{caption}
\usepackage{subcaption}
\usepackage{amsmath,stmaryrd,pict2e,picture}
\usepackage{float}
\usepackage{amsmath,amssymb,amsthm}
\usepackage{mathrsfs}
\usepackage{booktabs}
\usepackage{algorithm}
\usepackage{algpseudocode}
\usepackage{color, soul}
\usepackage{multirow}

\title{Angular Sector-Based Sparse Array Design for Adaptive Beamforming Using Deep Learning}

\author{
\IEEEauthorblockN{John~Kobak, Ethan~Atiyeh, and Syed~A.~Hamza}
\IEEEauthorblockA{
School of Engineering, Widener University, Chester, PA, USA\\
Emails: \{jjkobak, egatiyeh, shamza\}@widener.edu}
\thanks{This work was funded by the National Science Foundation (NSF) Award \#2347220.}
}

\begin{document}
\maketitle
\begin{abstract}
 Efficient sparse array reconfigurability is essential for cognitive sensing in dynamic radio frequency environments, where rapid interference variations require both adaptability and stability. This work presents a framework for designing sparse arrays optimized over broad angular sectors, enabling near-optimal beamforming that maximizes the signal-to-interference-plus-noise ratio (SINR) across a range of interferer angles. Full data correlation matrices are computed for candidate configurations, and an angular-sector–based class reduction strategy is applied to merge adjacent sectors dominated by the same configuration, resulting in 56 representative classes. Controlled up- and down-sampling produce four dataset variants involving, high and low sample count, balanced and unbalanced datasets, to systematically evaluate the effects of dataset size and class distribution on neural network performance. A lightweight convolutional neural network (CNN) and a deeper ResNet-50 architecture are trained and evaluated using these datasets. Results demonstrate high classification accuracy, with ResNet-50 achieving up to 97.3$\%$, while SINR deviations remain below 1$\%$ for most classes and below 5$\%$ even for challenging interference angles near broadside. The proposed approach enables robust sparse array selection, maintains strong SINR performance, reduces unnecessary reconfigurations, and provides an effective framework for real-time cognitive sensing and adaptive interference mitigation.
\end{abstract}

%\begin{keywords}
%Reconfigurable sparse arrays, CNN, cognitive sensing.
%\end{keywords}

\section{Introduction}
Maximizing signal-to-interference-plus-noise ratio (MaxSINR) in array processing often requires non-uniformly spaced sparse arrays, where inter-element spacing varies across the array \cite{6774934, 9257073, 9266359, 8061016, 9114655}. Sparse arrays provide enhanced degrees of freedom, with beamforming performance determined not only by array weights but also by the precise placement of array elements, unlike uniform linear arrays (ULAs) where beamforming depends solely on the weights \cite{HAMZA2020102678, 8378759, 8682266}.

In dynamic RF environments, source characteristics such as location, number, and power can change rapidly, necessitating responsive sparse array reconfigurability. Computing optimal array configurations and beamformer weights under these conditions is challenging, particularly within the Perception-Action Cycle (PAC) required for cognitive sensing \cite{10038907}.
Recent advances in convex optimization \cite{4663892} and fast machine learning methods \cite{articlelecun1, articlelecun3, inproceedingslecun2} have reduced computational complexity, enabling sparse arrays to serve as core components in adaptive, reconfigurable systems \cite{1139138, 9764243, 5456168, 7012090, 7444122, 9352510, 6714077, 6031934, 9746030}. Deep learning, particularly convolutional neural networks (CNNs), shifts much of the optimization burden to an offline training phase, where $N$ antennas are selected from a grid of $M$ possible locations to maximize array aperture. The trained model predicts active antenna locations, and an optimization algorithm subsequently computes beamformer weights to achieve MaxSINR \cite{10149410, 10818605, 10548736, 10570503, doi:https://doi.org/10.1002/9781394191048.ch7}.

Practical deployment also requires minimizing excessive antenna switching. As interference directions-of-arrival (DOAs) change, frequent reconfiguration may be impractical. While binary switching can simplify hardware \cite{10190882, 10094662}, maintaining sparse array configurations fixed across contiguous angular sectors until Signal-to-interference-plus-noise ratio (SINR) drops below a threshold is more effective. This approach reduces unnecessary reconfigurations, enhances computational efficiency, and improves classification performance by optimizing array behavior over broad angular regions rather than individual angles.

In this paper, datasets are generated by sweeping a single interferer across the field of view (FoV) while the desired signal remains at broadside. Full spatial correlation matrices are computed for each configuration under varying interference-to-noise ratios. Interference angles are sampled uniformly at 1° increments to ensure consistent coverage and prevent artificial clustering. Sparse array selection is performed across  candidate sparse array configurations. To reduce classification complexity while preserving near-optimal SINR, an angular-sector–based class reduction strategy merges adjacent sectors dominated by a single configuration, producing 56 representative classes. Controlled up-sampling and down-sampling then generate four dataset variants, high and low sample count, balanced and unbalanced, allowing systematic analysis of the effects of dataset size and class imbalance on neural network performance.

The resulting design approach ensures that the sparse arrays maintain strong SINR performance across the FoV while limiting the number of required reconfigurations. Evaluations of a lightweight CNN model and a deeper ResNet-50 architecture demonstrate high classification accuracy across all dataset variants, with ResNet-50 achieving up to 97.3$\%$ accuracy in the low-sample-count balanced dataset. SINR performance remains near-optimal, with average deviations below 1$\%$ for most classes and maximum deviations of only about 5$\%$ near broadside interference. Training efficiency and robustness of both networks indicate that the proposed approach is practical for real-time cognitive sensing, providing a scalable framework for adaptive sparse array deployment and effective interference mitigation in dynamic RF environments.

The paper is organized as follows: Section II formulates the problem; Section III details the CNN architectures and simulation setup; Section IV presents design examples and performance analysis; and Section V concludes the work.

%We note that while the spatial shifts and reflections about the array center points yield configurations with identical SINR, they do not generate identical configurations from a communications channel perspective []. 
%%%%%%%%%%%%%%%%%%%%%%%%%%%%%%%%%%%%%%%%%
\section{Problem Formulation} \label{Problem Formulation}

Consider  a desired  source and $L$ independent interfering  sources whose signals impinge  on a uniform linear array (ULA) with $M$ antennas. The baseband data received at the array at time $k$ is  given by, \vspace{-3mm}
\begin {equation} \label{a}
\mathbf{x}(k)=   \alpha(k) \mathbf{s}( \theta)  + \sum_{l=1}^{L} \beta _l(k) \mathbf{i}( \theta_l)  + \mathbf{v}(k), 
\end {equation}
where ($\alpha (n)$, $\beta _l(n))$ $\in \mathbb{C}$  are the complex amplitudes of the incoming baseband signals, $\mathbf{v}(n)$ $\in \mathbb{C}^M$ is additive Gaussian noise  with  variance   $\sigma_v^2$, and ($\mathbf{s} ({\theta})$, $\mathbf{i} ({\theta_l})$) $\in \mathbb{C}^M$ are  the  respective steering vectors corresponding to the directions of arrival, $\theta$ and $\theta_l$, of the desired source and $l$th interference, and are defined as,  \vspace{-2mm}
\begin {eqnarray}  \label{b}
\mathbf{s} ({\theta})&=&[1 \,  \, e^{j (2 \pi / \lambda) d \text{cos}(\theta)  } \,  ...\ e^{j (2 \pi / \lambda) d (M-1) \text{cos}(\theta)  }]^\text{T} \nonumber \\
\mathbf{i} ({\theta_l})&=&[1 \,  \,  e^{j (2 \pi / \lambda) d \text{cos}(\theta_l)  } \,  ... \, e^{j (2 \pi / \lambda) d (M-1) \text{cos}(\theta_l)  }]^\text{T}
\end {eqnarray}
Here, $d$ is the inter-element spacing and the superscript `$\text{T}$' denotes matrix transpose.   %\cite{trove.nla.gov.au/work/15617720}. 
The elements of $\mathbf{x}(k)$  are combined linearly by the $M$-sensor beamformer that strives to maximize the output SINR. The output signal $y(k)$ of the optimum beamformer for MaxSINR is given by \cite{1223538}, 
\begin {equation}  \label{c}
y(k) = \mathbf{w}_o^H \mathbf{x}(k), 
\end {equation} 
where the superscript `$H$' denotes Hermitian operation and $\mathbf{w}_o$ is the optimum weight vector resulting in the optimum output SINR$_o$, \vspace{-4mm}
\begin{equation}  \label{c2}
\text{SINR}_o=\frac {\mathbf{w}_o^H \mathbf{R}_s \mathbf{w}_o} { \mathbf{w}_o^H \mathbf{R}_{s^{'}} \mathbf{w}_o}. 
\end{equation}
For statistically independent signals, the desired source correlation matrix is $\mathbf{R}_s= \sigma^2 \mathbf{s}( \theta)\mathbf{s}^H( \theta)$, where $ \sigma^2 =E\{\alpha (k)\alpha ^H(k)\}$. Likewise,  the  interference and noise correlation matrix, $\mathbf{R}_{s^{'}}= \sum_{l=1}^{L} (\sigma^2_l \mathbf{i}( \theta_l)\mathbf{i}^H( \theta_l)$) + $\sigma_t^2\mathbf{I}_{M\times M}$, with $ \sigma^2_l =E\{\beta _l(k)\beta_l^H(k)\}$ being the power of the $l$th interfering source. 
In order to maximize the SINR expression in (\ref{c2}), we constrain the numerator and minimize the denominator as \vspace{-2mm}
\begin{equation} \label{d}
\begin{aligned}
\underset{\mathbf{w} \in \mathbb{C}^M}{\text{minimize}} & \quad   \mathbf{w}^H\mathbf{R}_{s^{'}}\mathbf{w},\\
\text{s.t.} & \quad     \mathbf{ w}^H\mathbf{R}_{s}\mathbf{ w}=1.
\end{aligned} 
\end{equation}
The problem in (\ref{d}) can be rewritten by replacing $\mathbf{R}_{s^{'}}$ with the received data covariance matrix, $\mathbf{R_{xx}}=\mathbf{R}_s+ \mathbf{R}_{s^{'}}$ \cite{1223538},\vspace{-1mm}
\begin{equation} \label{e}
\begin{aligned}
\underset{\mathbf{w} \in \mathbb{C}^M }{\text{minimize}} & \quad   \mathbf{ w}^H\mathbf{R_{xx}}\mathbf{ w},\\
\text{s.t.} & \quad     \mathbf{ w}^H\mathbf{R}_{s}\mathbf{ w} \geq 1.
\end{aligned} 
\end{equation}

To bring aperture sparsity into optimum beamformer design, the  constrained optimization (\ref{e}) can be reformulated  by incorporating  sparsity enhancing $l_1$-norm as follows; 
\begin{equation} \label{a2}
\begin{aligned}
\underset{\mathbf{{\bf w} \in {\mathcal C}}^{M}}{\text{minimize}} & \quad  \mathbf{ w}^H\mathbf{R_{xx}}\mathbf{w},\\
\text{s.t.} & \quad   \mathbf{ w}^H\mathbf{R}_{s}\mathbf{ w}\geq1 \   \text{and} \ ||\mathbf{w}||_1=ND
\end{aligned}  
\end{equation}
 Here, $||.||_1$ is the $l_1$ norm, which  ensures  beamforming with the cardinality of the weight vector $\mathbf{w}$  equal to the number of available RF chains, $N$.  We assume a case for  full switching networks where each  antenna is connected with all RF chains. This results in a combinatorial optimization problem and can be solved  by enumerating over all possible sensor locations or employing  data-dependent convex relaxation algorithms  \cite{6477161}. These algorithms can successfully yield sparse arrays ensuring only $N$ active sensors, but they lack the ability to configure the array design in real time due to high computational complexity, especially in applications involving rapidly changing environments. 
 
 %Different binary switching strategies are possible depending on the choice of antennas that are paired together. 

 In this paper, DL approaches are developed by learning the  sparse array configuration through enumeration. The problem in (\ref{e}) is solved for all possible sparse array configurations to arrive at sparse array which achieves MaxSINR performance. The DL network is then trained to provide a direct mapping of the sensor data correlations  to the optimum switched sparse array configuration for a given `look' direction. It is noted that the DL network only learns the optimum array configuration while beamformer weights are obtained by solving (6) for the specific array configuration obtained through DL. 
 %%%%%%%%%%%%%%%%%%%%%%%%%%%%%%%%%%%%%%%
\section{Simulation Setup} \label{Simulations} 
We consider a scenario with $M=12$ possible antenna locations and $N=6$ RF chains, resulting in 924 potential sparse array configurations. To reduce system complexity and facilitate neural network training, a subset of 56 representative configurations is selected from these 924 possibilities, as explained below.

\subsection{Dataset Generation}
\label{sec:data_gen}

For data generation, full array data correlation matrices are computed for each configuration. The desired signal is fixed at broadside ($90^\circ$) with a signal-to-noise ratio (SNR) of 0~dB, while the interference is varied across the  FoV with interference-to-noise ratio (INR) values ranging from 15–20~dB. Each $1^\circ$ angular bin in the FoV, spanning $10^\circ$ to $90^\circ$, contains multiple interference angle samples to represent different realizations of the interferer for each configuration. For example, a total of 8000 samples would allocate 100 samples per degree across this angular range.

To assign sparse array configurations to interferer angles, the FoV is divided into 80 initial angular bins. For each bin, the configuration that most frequently achieves near-optimal SINR is selected. In the upper portion of the FoV ($80^\circ$–$90^\circ$), a single configuration dominates multiple adjacent bins, prompting the merging of these bins into larger sectors and reducing the total number of distinct bins to 72. Because of repeating optimum arrays, in a few bins, we have 56 unique sector-based configurations that form the labels used for neural network training and testing.
The datasets are partitioned into training and testing subsets using an 80/20 split, with random samples drawn from each angular bin so that both sets include a representative range of closely spaced angles. This ensures that the network is tested on angles similar to those seen during training, avoiding unrealistic separation of adjacent interferer angles. 

\begin{figure}[t!]
    \centering
    \subfloat[]{%
      \includegraphics[clip,width=0.70\columnwidth]{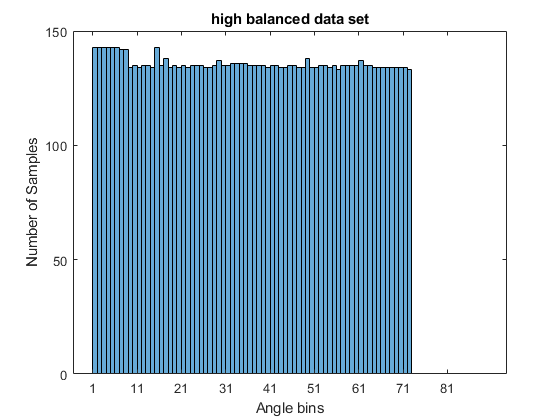}%
      \label{fig:HB}
    }

    \subfloat[]{%
      \includegraphics[clip,width=0.70\columnwidth]{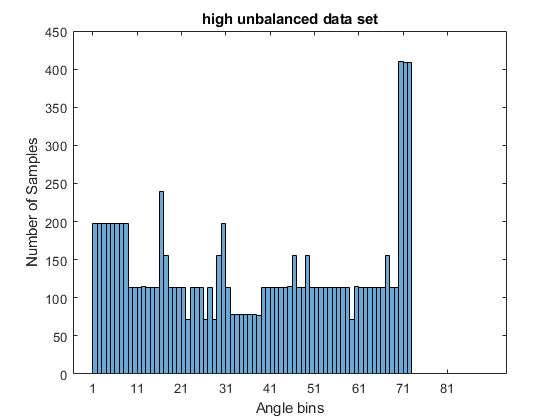}%
      \label{fig:HU}
    }

    \subfloat[]{%
      \includegraphics[clip,width=0.70\columnwidth]{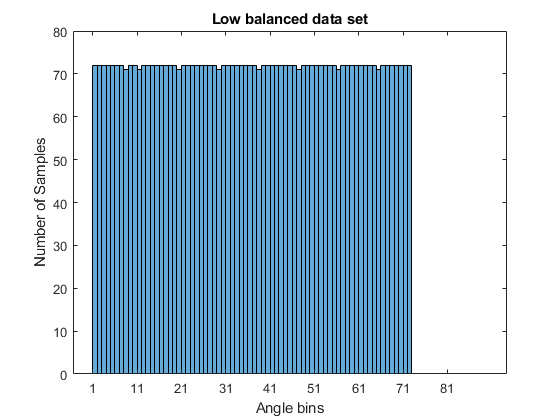}%
      \label{fig:LB}
    }
    
    \subfloat[]{%
      \includegraphics[clip,width=0.70\columnwidth]{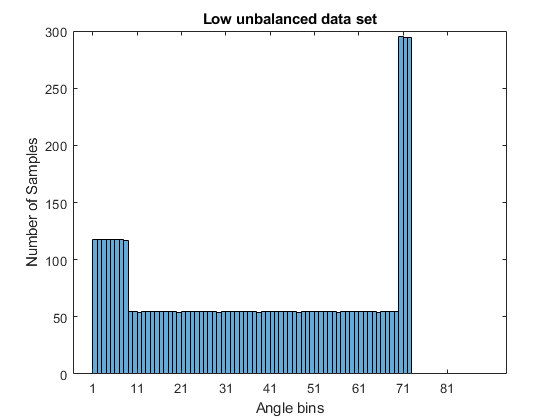}%
      \label{fig:LU}
    }
    \caption{The four data sets: High Balanced, High Unbalanced, Low Balanced, Low Unbalanced}
    \label{Histograms of data}
\end{figure}

While this binning procedure reduces label complexity and improves task learnability, it does not entirely eliminate class imbalance, as certain configurations remain more prevalent due 
\begin{table}[t]
\centering
\resizebox{\linewidth}{0.35 in}{%
\begin{tabular}{l|c|c|c|c}
\toprule
& \textbf{High\_Balanced} & \textbf{High\_Unbalanced} & \textbf{Low\_Balanced} & \textbf{Low\_Unbalanced} \\
\midrule
\# Samples     & 9778 & 9758 & 5176 & 5174 \\
\midrule
Samples/Bin    & 135  & N/A  & 72   & N/A  \\
\bottomrule
\end{tabular}
}
\caption{Dataset sample counts and samples-per-bin for each configuration.}
\label{tab:dataset_counts}
\end{table}
to the underlying SINR geometry. To systematically evaluate the impact of sample quantity and bin balance on neural network performance, four dataset variants are constructed, as  detailed in Table \ref{tab:dataset_counts} and illustrated in Fig. \ref{Histograms of data}. The first is a high-sample-count unbalanced dataset that preserves the natural bin distribution dictated by SINR-optimal configurations. From this baseline, a high-sample-count balanced dataset is created using a combination of up-sampling for minority bins and selective down-sampling for majority bins to achieve uniform bin representation while maintaining the overall dataset size. For data-limited regimes, the total number of interference angle samples per bin is reduced to produce a low-sample-count unbalanced dataset, retaining the original imbalance characteristics but with fewer examples per class. A corresponding low-sample-count balanced dataset is then generated by equalizing bin populations. These four datasets, differing only in sample density and bin distribution, enable isolation of the effects of dataset size versus bin balance on neural network generalization.
Once the labeled correlation matrices are generated, both the reduced-complexity CNN model and the deeper ResNet-50 architecture \cite{he2016deep} are trained and evaluated using all four dataset variants. Each correlation matrix, represented by its real, imaginary, and phase components, serves as the network input, while the target output is a one-hot encoded vector corresponding to one of the 56 sector-based sparse array configurations.

\subsection{Network Topology}

\begin{figure}[t!]
    \centering
    \includegraphics[width=0.8\linewidth,height=1.7 in]{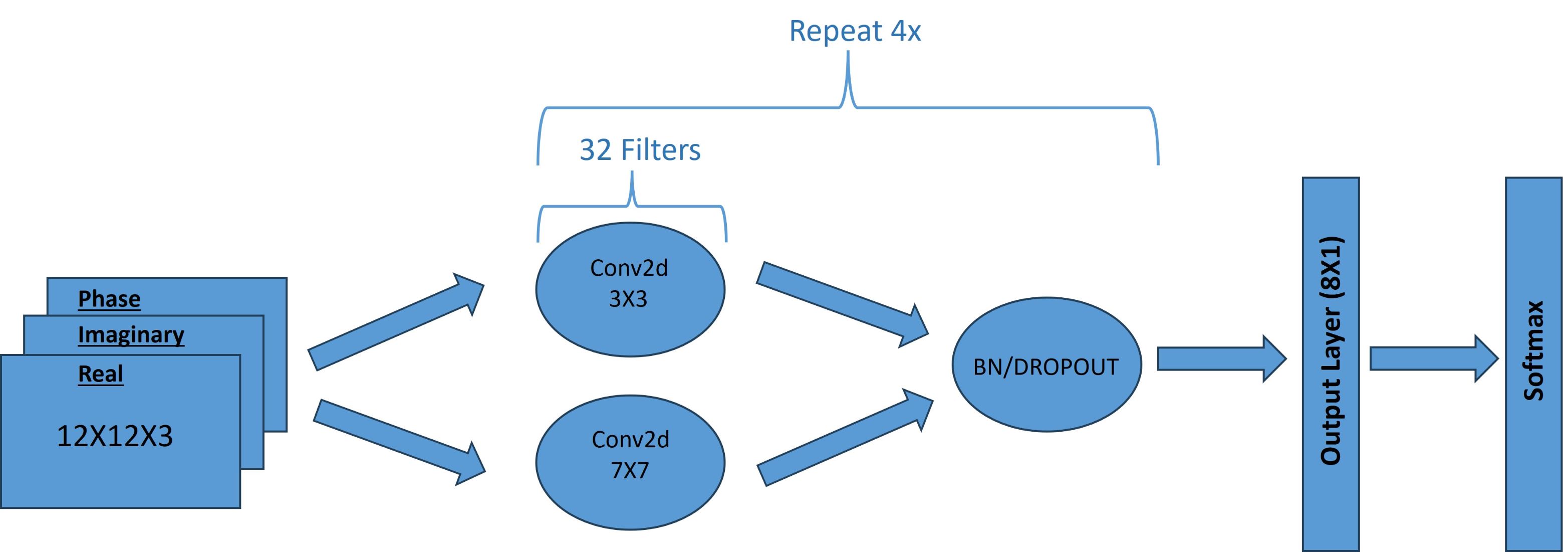}
    \caption{CNN Architecture} 
\label{CNN_diagram}
\end{figure}

 The employed CNN is shown in  Fig. \ref{CNN_diagram}. The input is of size $3 \times M\times M$ and comprises the real, imaginary parts and phase  of the $M \times M$ correlation matrix. It is noted that  sparse array design can only have a few active sensors at a time which makes it difficult to  obtain full data  correlation values corresponding to the inactive sensor locations as required by the DL approaches. However, for the scope of this paper, we  assume that  estimates of all the  correlation lags corresponding to the  full  aperture array are available. This can typically be achieved by employing  a  low rank matrix completion  strategy or to sequentially estimate the missing data correlation values over different subarrays which are configured through antenna switching.  The output layer is size 56 and represents the possible sparse array configurations.  The neural network architecture employed is shown in Fig. 2 and it implements 32 parallel filters, in each layer, of size $3 \times 3$ and $7 \times 7$ and  incorporating dropout and batch normalization techniques. The network output is a one-hot encoded vector. The ReLU activation function is used for all layers except the output layer, which uses a softmax function. 
 
The network training utilized a batch size of 128 with a binary cross entropy loss function and ran for 10 epochs. The Adam optimizer \cite{ref:adam} uses a cosine annealing learning rate \cite{Loshchilov2016SGDR} using a schedule, 
\begin{equation}
\eta(t) = \eta_0 \left[
\alpha_l (1 - \alpha_l)\,
\frac{1}{2}\left(1 + \cos\left(\pi \frac{t}{T}\right)\right)
\right]
\label{eq:cosine_decay}
\end{equation}
where $\eta_t$ is the learning rate at step t, $\eta_0$ is the initial learning rate, $T$ is the number of decay steps, and $\alpha_l$ is the final learning rate fraction. In this paper, the initial learning rate was $1\text{e}^{-4}$, 10,000 decay steps, and $\alpha_l$ = $1\text{e}^{-5}$. Early stopping implements after 25 epochs, based on the highest validation accuracy observed. 

The second model is a deeper ResNet-50 architecture using residual skip connections to enable higher-level feature learning. The learning rate for the ResNet is slightly higher with an initial rate of $3\text{e}^{-4}$ and a final evaluation learning rate of $1\text{e}^{-3}$. Early stopping was initiated after 6 epochs and both models are trained for ten epochs per run. %The network weights were then fine-tuned with four additional cycles of learning, starting with the network weights of the previous run after each iteration. 

Both models are trained by using a hybrid $K$-fold validation procedure intended to improve the networks performance. In standard $K$-fold validation, the dataset is divided into $K$ subsets. Both models are split into 5 subsets. Each subset serves once as the validation set, providing a more dependable estimate of the network's ability to generalize rather than a single split. With this hybrid approach, each fold did not start with random weights. Instead, it starts with the best weights from the previous fold. After completing all folds, the fold with the highest validation accuracy is chosen and its corresponding weights are used on the test set. This strategy accelerates convergence and keeps useful feature characterizations across the 5 folds. As a result, both networks improve overall generalization across all four datasets.

%\begin{figure}[t!]
%    \centering 
    %\includegraphics[width=3.08 in, height=1.75 in]{./fig/Histogram of test classes post added 100 degree.jpg}
 %   \caption{Histogram of test classes pre equalizing.}
%\label{fig:mlp_network}
%\end{figure}

%\begin{figure}[t!]
%    \centering 
    %\includegraphics[width=3.08 in, height=1.75 in]{./fig/Histogram of test classes post equalizing.jpg}
%    \caption{Histogram of test classes post equalizing.}
%\label{fig:mlp_network}
%\end{figure}

\section{Results}

Table \ref{tab: NN results} summarizes the classification accuracy achieved by the CNN and ResNet-50 architectures for the four dataset variants. The CNN attains accuracies of 91.1$\%$ for the high-sample balanced dataset, 91.4$\%$ for the high-sample unbalanced dataset, 94.6$\%$ for the low-sample balanced dataset, and 92.3$\%$ for the low-sample unbalanced dataset. ResNet-50 achieves higher accuracy across all cases, with 93.1$\%$ for high-sample balanced, 94.8$\%$ for high-sample unbalanced, 97.3$\%$ for low-sample balanced, and 97.1$\%$ for low-sample unbalanced.

\begin{table}[t]
\centering
\resizebox{\linewidth}{0.35 in}{%
\begin{tabular}{l|c|c|c|c}
\toprule
& \textbf{High\_Balanced} & \textbf{High\_Unbalanced} & \textbf{Low\_Balanced} & \textbf{Low\_Unbalanced} \\
\midrule
CNN     & 91.1 & 91.4 & 94.6 & 92.3 \\
\midrule
ResNet-50    & 93.1  & 94.8  & 97.3   & 97.1  \\
\bottomrule
\end{tabular}
}
\caption{Classification accuracy for the four datasets and CNN/ResNet}
\label{tab: NN results}
\end{table} 

These results show a consistent trend across both networks: low-sample datasets yield higher accuracy than their high-sample counterparts, and class balancing improves accuracy only in the low-sample regime. The higher accuracy observed in the low-sample datasets arises because the reduced number of samples produces more compact class clusters in the feature space. With fewer realizations per angular bin, the intra-class variability is significantly smaller, making the classification problem easier for both networks. In contrast, the high-sample datasets include many more interference realizations under varying INR conditions, which increases intra-class diversity and introduces additional overlap near sector boundaries, reducing classification accuracy.

\begin{figure}
    \centering
    \subfloat[]{%
      \includegraphics[clip,width=0.8\columnwidth]{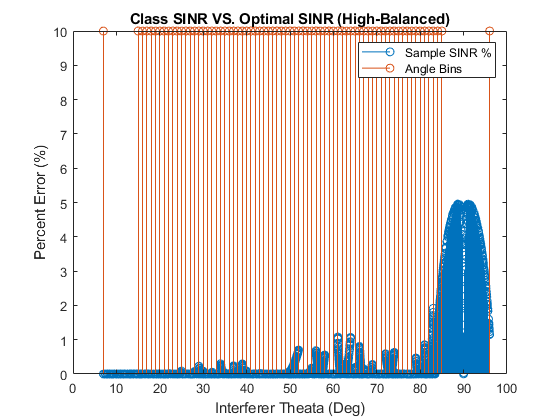}%
      \label{fig:HB}
    }

    \subfloat[]{%
      \includegraphics[clip,width=0.8\columnwidth]{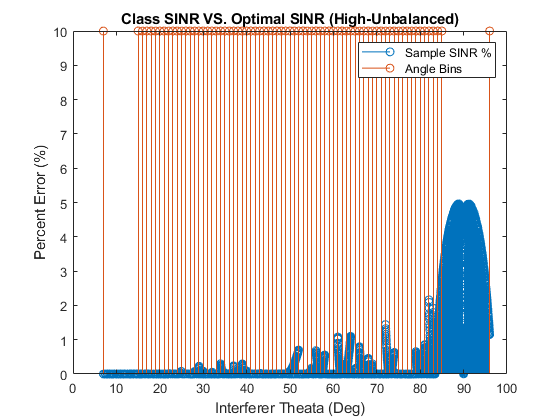}%
      \label{fig:HU}
    }

    \subfloat[]{%
      \includegraphics[clip,width=0.8\columnwidth]{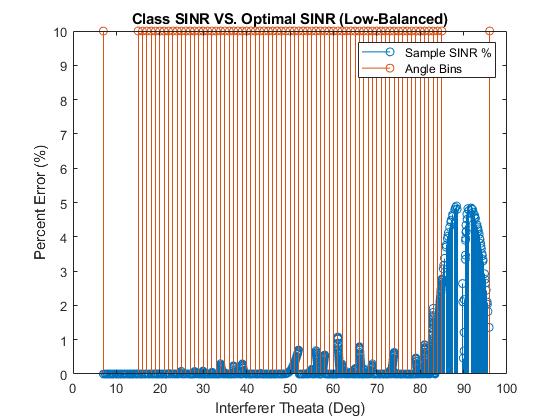}%
      \label{fig:LB}
    }
    
    \subfloat[]{%
      \includegraphics[clip,width=0.8\columnwidth]{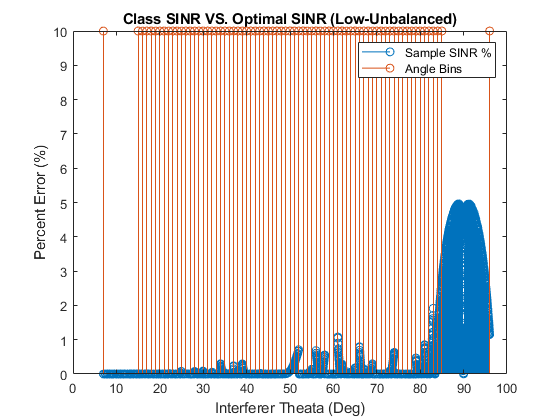}%
      \label{fig:LU}
    }
    \caption{The four datasets comparing the percent error between the chosen array configuration and the optimal across all bins}
    \label{SINR graphs}
\end{figure}

The effect of class balancing also differs between the two regimes. In the low-sample case, several classes contain very few examples, limiting the network’s ability to learn their distinguishing features. Balancing these datasets through controlled up-sampling improves representation for minority classes and leads to higher accuracy. However, in the high-sample datasets, majority classes already contain sufficient diversity. Down-sampling these classes or over-replicating minority classes disrupts the natural distribution dictated by the underlying SINR geometry, reducing the diversity of dominant classes and causing the network to overfit repeated samples. As a result, balancing slightly degrades performance in the high-sample regime.

Fig. \ref{SINR graphs} illustrates the SINR behavior across the 72 bins by reporting the percentage deviation from the optimal SINR. For most bins, the deviation remains below 1$\%$. As the interferer approaches broadside, where the desired source is located, the degradation increases, consistent with the increased difficulty of interference suppression near this region. Even in these cases, the maximum deviation is only about 5$\%$. Averaged over all samples in the dataset, the achieved SINR is 5.628 dB, demonstrating that the sector-based configuration selection preserves near-optimal performance across the FoV.

Together, these results confirm that sector-based class reduction, combined with careful dataset construction, enables reliable sparse-array configuration prediction while maintaining strong SINR performance. ResNet-50 consistently provides higher accuracy due to its deeper feature-extraction capability, while the lighter CNN remains competitive with significantly lower computational cost, making both models viable for real-time cognitive sensing applications depending on system constraints.

\section{Conclusion}
The paper considered a neural network approach for selecting sparse array configurations using array correlation data and angular-sector based class labels. Both a conventional CNN and a deeper ResNet-50 were trained and tested on datasets with different sample counts and class balance. The ResNet-50 consistently achieved higher classification accuracy, especially when sample sizes were small, showing the advantages of deeper networks in data-limited scenarios. Balancing the classes improved performance for small datasets but had less impact as the number of samples increased. The networks maintained near-optimal SINR across all samples with only small deviations for challenging interferer angles. These results show that neural networks can effectively identify sparse array configurations while preserving strong SINR performance and that careful dataset design is important for achieving reliable learning outcomes.
\balance

\bibliographystyle{IEEEtran}
\bibliography{references.bib}

@INPROCEEDINGS{10190882, 
  author={Hamza, Syed A. and Juretus, Kyle and Amin, Moeness G. and Ahmad, Fauzia},
  booktitle={2023 International Symposium on Signals, Circuits and Systems (ISSCS)}, 
  title={Deep Learning Sparse Array Design Considering Binary Switching and Missing Coarray Lags}, 
  year={2023},
  volume={},
  number={},
  pages={1-4},
  doi={10.1109/ISSCS58449.2023.10190882}}

@INPROCEEDINGS{10094662,
  author={Hamza, Syed A. and Juretus, Kyle and Amin, Moeness G. and Ahmad, Fauzia},
  booktitle={ICASSP 2023 - 2023 IEEE International Conference on Acoustics, Speech and Signal Processing (ICASSP)}, 
  title={Deep Learning Sparse Array Design Using Binary Switching Configurations}, 
  year={2023},
  volume={},
  number={},
  pages={1-5},
  doi={10.1109/ICASSP49357.2023.10094662}}

@INPROCEEDINGS{10038907,
  author={Hamza, S. A. and Amin, M. G. and Kirk, B. and Martone, A.},
  booktitle={International Conference on Radar Systems (RADAR 2022)}, 
  title={Sparse array reconfigurability for source identification and angle estimation in cognitive sensing}, 
  year={2022},
  volume={2022},
  number={},
  pages={83-88},
  doi={10.1049/icp.2022.2296}}

@INPROCEEDINGS{9764243,
  author={Amin, Moeness G. and Hamza, Syed A. and Kirk, Benjamin and Martone, Anthony},
  booktitle={2022 IEEE Radar Conference (RadarConf22)}, 
  title={Fast DOA Estimation Using Coarray Beamforming with Model Order Estimation}, 
  year={2022},
  volume={},
  number={},
  pages={1-6},
  doi={10.1109/RadarConf2248738.2022.9764243}}

@INPROCEEDINGS{9746030,
  author={Hamza, Syed A. and Amin, Moeness G. and Chalise, Batu K.},
  booktitle={ICASSP 2022 - 2022 IEEE International Conference on Acoustics, Speech and Signal Processing (ICASSP)}, 
  title={Phase-Only Reconfigurable Sparse Array Beamforming Using Deep Learning}, 
  year={2022},
  volume={},
  number={},
  pages={4913-4917},
  doi={10.1109/ICASSP43922.2022.9746030}}

@ARTICLE{10149410,
  author={Hamza, Syed A. and Amin, Moeness G.},
  journal={IEEE Transactions on Aerospace and Electronic Systems}, 
  title={Sparse Array Design for Optimum Beamforming Using Deep Learning}, 
  year={2024},
  volume={60},
  number={1},
  pages={133-144},
  doi={10.1109/TAES.2023.3285201}}

@ARTICLE{10818605,
  author={Juretus, Kyle and Amin, Moeness G. and Hamza, Syed A.},
  journal={IEEE Transactions on Aerospace and Electronic Systems}, 
  title={Deep Learning of the Sparse Array Configurations in Optimum Beamforming}, 
  year={2024},
  volume={},
  number={},
  pages={1-12},
  doi={10.1109/TAES.2024.3523276}}

@INPROCEEDINGS{10548736,
  author={Amin, Moeness G. and Hamza, Syed A. and Juretus, Kyle},
  booktitle={2024 IEEE Radar Conference (RadarConf24)}, 
  title={Sparse Array Configuration Analysis and Deep Learning Classifications for Beamfornming}, 
  year={2024},
  volume={},
  number={},
  pages={1-6},
  doi={10.1109/RadarConf2458775.2024.10548736}}

@INPROCEEDINGS{10570503,
  author={Hamza, Syed A. and Amin, Moeness G. and Juretus, Kyle},
  booktitle={2024 IEEE Wireless Communications and Networking Conference (WCNC)}, 
  title={On the Roles of Sparse Array Configuration and Weights in Optimum Beamforming}, 
  year={2024},
  volume={},
  number={},
  pages={1-6},
  doi={10.1109/WCNC57260.2024.10570503}}

@inbook{doi:https://doi.org/10.1002/9781394191048.ch7,
author = {Hamza, Syed A. and Juretus, Kyle and Amin, Moeness G.},
publisher = {John Wiley \& Sons, Ltd},
isbn = {9781394191048},
title = {Sparse Array Design for Optimum Beamforming Using Deep Learning},
booktitle = {Sparse Arrays for Radar, Sonar, and Communications},
chapter = {7},
pages = {215-250},
doi = {https://doi.org/10.1002/9781394191048.ch7},
url = {https://onlinelibrary.wiley.com/doi/abs/10.1002/9781394191048.ch7},
eprint = {https://onlinelibrary.wiley.com/doi/pdf/10.1002/9781394191048.ch7},
year = {2024}
}

@misc{ref:adam,
  doi = {10.48550/ARXIV.1412.6980},
  url = {https://arxiv.org/abs/1412.6980},
  author = {D. P. Kingma and J. Ba},
  title = {Adam: A Method for Stochastic Optimization},
  publisher = {arXiv},
  year = {2014},
}

@ARTICLE{9257073,
  author={S. A. {Hamza} and M. G. {Amin}},
  journal={IEEE Transactions on Aerospace and Electronic Systems}, 
  title={Sparse Array Beamforming Design for Wideband Signal Models}, 
  year={2020},
  volume={},
  number={},
  pages={1-1},
  doi={10.1109/TAES.2020.3037409}}

@INPROCEEDINGS{9266359,
  author={S. A. {Hamza} and M. G. {Amin}},
  booktitle={2020 IEEE Radar Conference (RadarConf20)}, 
  title={Learning Sparse Array Capon Beamformer Design Using Deep Learning Approach}, 
  year={2020},
  volume={},
  number={},
  pages={1-5},
  doi={10.1109/RadarConf2043947.2020.9266359}}

@INPROCEEDINGS{9114655,
  author={S. A. {Hamza} and M. G. {Amin}},
  booktitle={2020 IEEE International Radar Conference (RADAR)}, 
  title={Sparse Array Design for Transmit Beamforming}, 
  year={2020},
  volume={},
  number={},
  pages={560-565},
  doi={10.1109/RADAR42522.2020.9114655}}

@ARTICLE{9352510,
  author={Rajamäki, Robin and Koivunen, Visa},
  journal={IEEE Transactions on Signal Processing}, 
  title={Sparse Symmetric Linear Arrays With Low Redundancy and a Contiguous Sum Co-Array}, 
  year={2021},
  volume={69},
  number={},
  pages={1697--1712},
  doi={10.1109/TSP.2021.3057982}}

@ARTICLE{7444122,
  author={Amin, Moeness G. and Wang, Xiangrong and Zhang, Yimin D. and Ahmad, Fauzia and Aboutanios, Elias},
  journal={Proceedings of the IEEE}, 
  title={Sparse Arrays and Sampling for Interference Mitigation and {DOA} Estimation in {GNSS}}, 
  year={2016},
  volume={104},
  number={6},
  pages={1302-1317},
  doi={10.1109/JPROC.2016.2531582}}

@article{articlelecun1,
author = {LeCun, Yann and Bengio, Y. and Hinton, Geoffrey},
year = {2015},
month = {05},
pages = {436-44},
title = {Deep Learning},
volume = {521},
journal = {Nature},
doi = {10.1038/nature14539}
}

@inproceedings{inproceedingslecun2,
author = {Deng, li and Li, Jinyu and Huang, Jui-Ting and Yao, Kaisheng and Yu, Dong and Seide, Frank and Seltzer, Michael and Zweig, Geoff and He, Xiaodong and Williams, Jason and Gong, Yifan and Acero, Alex},
year = {2013},
month = {10},
pages = {8604-8608},
title = {Recent advances in deep learning for speech research at Microsoft},
journal = {Acoustics, Speech, and Signal Processing, 1988. ICASSP-88., 1988 International Conference on},
doi = {10.1109/ICASSP.2013.6639345}
}

@article{articlelecun3,
author = {Krizhevsky, Alex and Sutskever, Ilya and Hinton, Geoffrey},
year = {2012},
month = {01},
pages = {},
title = {ImageNet Classification with Deep Convolutional Neural Networks},
volume = {25},
journal = {Neural Information Processing Systems},
doi = {10.1145/3065386}
}

@article{HAMZA2020102678,
title = "Sparse array design for maximizing the signal-to-interference-plus-noise-ratio by matrix completion",
journal = "Digital Signal Processing",
pages = "102678",
year = "2020",
issn = "1051-2004",
doi = "https://doi.org/10.1016/j.dsp.2020.102678",
author = "Syed A. Hamza and Moeness G. Amin"
}

@INPROCEEDINGS{8378759,
author={S. A. {Hamza} and M. G. {Amin} and G. {Fabrizio}},
booktitle={2018 IEEE Radar Conference (RadarConf18)},
title={Optimum sparse array beamforming for general rank signal models},
year={2018},
volume={},
number={},
pages={1343-1347},
doi={10.1109/RADAR.2018.8378759},
ISSN={},
month={April},}

@INPROCEEDINGS{8682266, 
author={S. A. {Hamza} and M. G. {Amin}}, 
booktitle={ICASSP 2019 - 2019 IEEE International Conference on Acoustics, Speech and Signal Processing (ICASSP)}, 
title={Hybrid Sparse Array Design for Under-determined Models}, 
year={2019}, 
volume={}, 
number={}, 
pages={4180-4184}, 
doi={10.1109/ICASSP.2019.8682266}, 
ISSN={2379-190X}, 
month={May},}

@ARTICLE{6031934, 
author={H. Godrich and A. P. Petropulu and H. V. Poor}, 
journal={IEEE Transactions on Signal Processing}, 
title={Sensor Selection in Distributed Multiple-Radar Architectures for Localization: {A} Knapsack Problem Formulation}, 
year={2012}, 
volume={60}, 
number={1}, 
pages={247-260}, 
doi={10.1109/TSP.2011.2170170}, 
ISSN={1053-587X}, 
month={Jan},}

@ARTICLE{6477161, 
 author={O.\ Mehanna and N.\ D.\ Sidiropoulos and G.\ B.\ Giannakis}, 
 journal={IEEE Transactions on Signal Processing}, 
 title={Joint Multicast Beamforming and Antenna Selection}, 
 year={2013}, 
 volume={61}, 
 number={10}, 
 pages={2660-2674}, 
 month={May},}

@ARTICLE{4663892, 
author={S. Joshi and S. Boyd}, 
journal={IEEE Transactions on Signal Processing}, 
title={Sensor Selection via Convex Optimization}, 
year={2009}, 
volume={57}, 
number={2}, 
pages={451-462}, 
doi={10.1109/TSP.2008.2007095}, 
ISSN={1053-587X}, 
month={Feb},}

@ARTICLE{1139138, 
author={A. Moffet}, 
journal={IEEE Transactions on Antennas and Propagation}, 
title={Minimum-redundancy linear arrays}, 
year={1968}, 
volume={16}, 
number={2}, 
pages={172-175}, 
doi={10.1109/TAP.1968.1139138}, 
ISSN={0018-926X}, 
month={March},}

@ARTICLE{6774934, 
 author={X.\ Wang and E.\ Aboutanios and M.\ Trinkle and M.\ G.\ Amin}, 
 journal={IEEE Transactions on Signal Processing}, 
 title={Reconfigurable Adaptive Array Beamforming by Antenna Selection}, 
 year={2014}, 
 volume={62}, 
 number={9}, 
 pages={2385-2396}, 
 month={May},}

@INPROCEEDINGS{6714077, 
 author={V.\ Roy and S.\ P.\ Chepuri and G.\ Leus}, 
 booktitle={2013 5th IEEE International Workshop on Computational Advances in Multi-Sensor Adaptive Processing (CAMSAP)}, 
 title={Sparsity-enforcing sensor selection for {DOA} estimation}, 
 year={2013}, 
 pages={340-343}, 
 month={Dec.},}

@ARTICLE{5456168, 
 author={P.\ Pal and P.\ P.\ Vaidyanathan}, 
 journal={IEEE Transactions on Signal Processing}, 
 title={Nested Arrays: A Novel Approach to Array Processing With Enhanced Degrees of Freedom}, 
 year={2010}, 
 volume={58}, 
 number={8}, 
 pages={4167-4181},
 month={Aug.},}

@ARTICLE{7012090, 
 author={S.\ Qin and Y.\ D.\ Zhang and M.\ G.\ Amin}, 
 journal={IEEE Transactions on Signal Processing}, 
 title={Generalized Coprime Array Configurations for Direction-of-Arrival Estimation}, 
 year={2015}, 
 volume={63}, 
 number={6}, 
 pages={1377-1390}, 
 month={March},}

@ARTICLE{8061016, 
 author={X. Wang and M. Amin and X. Cao}, 
 journal={IEEE Transactions on Signal Processing}, 
 title={Analysis and Design of Optimum Sparse Array Configurations for Adaptive Beamforming}, 
 year={2017}, 
 volume={PP}, 
 number={99}, 
 pages={1-1}, 
 doi={10.1109/TSP.2017.2760279}, 
 ISSN={1053-587X}, 
 month={},}

@ARTICLE{1223538, 
 author={S.\ Shahbazpanahi and A.\ B.\ Gershman and Zhi-Quan Luo and Kon Max Wong}, 
 journal={IEEE Transactions on Signal Processing}, 
 title={Robust adaptive beamforming for general-rank signal models}, 
 year={2003}, 
 volume={51}, 
 number={9}, 
 pages={2257-2269}, 
 month={Sept.},}

@INPROCEEDINGS{he2016deep,
  title={Deep Residual Learning for Image Recognition},
  author={He, Kaiming and Zhang, Xiangyu and Ren, Shaoqing and Sun, Jian},
  booktitle={Proceedings of the IEEE Conference on Computer Vision and Pattern Recognition (CVPR)},
  pages={770--778},
  year={2016},}

@inproceedings{Loshchilov2016SGDR,
  title={SGDR: Stochastic Gradient Descent with Warm Restarts},
  author={Loshchilov, Ilya and Hutter, Frank},
  booktitle={International Conference on Learning Representations (ICLR) Workshops},
  year={2017},
  note={arXiv preprint arXiv:1608.03983}
}

\end{document}